\documentclass[final,3p,times,twocolumn]{elsarticle}

\usepackage{graphicx}

\usepackage{amssymb}

\usepackage[utf8]{inputenc}
\usepackage{textcomp}

\usepackage{booktabs}
\usepackage{rotating}
\usepackage{multirow}
\usepackage{caption}
\usepackage{subcaption}

\usepackage{color}
\usepackage[usenames,dvipsnames,svgnames,table]{xcolor}

\journal{Journal of Nuclear Materials }

\begin{document}

\begin{frontmatter}



\title{OKMC simulations of Fe-C systems under irradiation: sensitivity studies}


\author[helsinki]{V.~Jansson\corref{cor1}}
\ead{ville.b.c.jansson@gmail.com}
\author[sck]{L.~Malerba}

\cortext[cor1]{Corresponding author. Tel. +358 2941 50088.}

\address[helsinki]{Helsinki Institute of Physics and Department of Physics, P.O. Box 43 (Pehr
Kalms gata 2), FI-00014 {\sc University of Helsinki, Finland}}

\address[sck]{Institute of Nuclear Materials Science, SCK$\bullet$CEN, Boeretang
200, 2400
{\sc Mol, Belgium}}

\begin{abstract}

This paper continues our previous work on a nanostructural evolution model for Fe-C alloys under irradiation, using Object Kinetic Monte Carlo modeling techniques. We here present a number of sensitivity studies of parameters of the model, such as the carbon content in the material, represented by generic traps for point defects, the importance of traps, the size dependence of traps and the effect of the dose rate. 
\end{abstract}

\begin{keyword}
Fe-C alloys \sep Object Kinetic Monte Carlo

\end{keyword}

\end{frontmatter}

\frenchspacing


\section{Introduction}

Iron (Fe) and carbon (C) are the basic components of any steel and particularly
of the low alloy banitic steels used in the reactor pressure vessels (RPV) of most commercial nuclear power plants. Neutron irradiation induces the creation of point defects, \textit{i.e.} self-interstitial atoms (SIA) and vacancies (V), isolated or in clusters. These, over the life time of the RPV, will have significant impact on the integrity of the material. Computer simulation models are an effective way to understand these kind of degradation processes and a first step in building a physical model for RPV steels is to address the Fe-C system.

In \cite{jansson2013simulation}, we presented an Object Kinetic Monte Carlo
(OKMC) model to describe the nanostructural evolution under irradiation in Fe-C alloys at temperatures $<$350 K, as well as post-irradiation annealing at temperatures up to 700 K. In the model, the effect of C was introduced by means of effective traps for vacancy and SIA type defects. The parameters used were strictly derived from physical considerations or atomistic calculations, or determined by values that reflected the reference experimental conditions. Only one parameter was used to guarantee the best reproduction possible of experimental data via fitting. In this paper, we examine the sensitivity of the model to the variation of this, as well as other parameters, such as the concentration of carbon in terms of traps, other features of the traps, and an environmental parameter such as the dose-rate.

The simulation results are compared for convenience with the low temperature ($<$350 K) irradiation experiment by Eldrup, Singh and Zinkle
\cite{zinkle2006microstructure,eldrup2002dose}, which was also used as a
reference experiment in \cite{jansson2013simulation}. This is one of the most
complete irradiation experiments for pure iron because both the
evolution with dose of vacancies and SIA clusters has been traced, using respectively positron annihilation spectroscopy (PAS) and transmission electron
microscopy (TEM). We have also included experimental data from
other TEM studies from similar experiments, such as
\cite{singh1999effects,bryner1966study,robertson1982low,horton1982tem,
takeyama1981,eyre1965electron}. A more thorough overview of the available irradiation and
annealing experiments in Fe-C will be published in a separate paper \cite{malerba2011review}.

This paper is organized as follows: In Sec. \ref{sec:methods} our computational method is briefly overviewed. In Sec. \ref{sec:traps}, we study the nanostructure evolution when no traps are present. In Sec. \ref{sec:C_density}, we study the effect of the trap concentration, which correlates with the carbon content. In Sec. \ref{sec:threshold} and \ref{sec:etlarge}, the sensitivity of variables
giving the size dependence of the SIA traps are studied. 
In Sec. \ref{sec:flux}, we study the model's sensitivity to the dose rate. Finally, we
discuss the results and present our conclusions in Sec. \ref{sec:discussion}.

\section{Computation method}\label{sec:methods}

For our OKMC simulations, we use the code {\sc LAKIMOCA}, thoroughly described in \cite{domain2004simulation}. Our methodology is described in \cite{jansson2013simulation}, of which this paper is a
direct follow-up. Here we point out only the main features of the method.

Following our work in \cite{jansson2013simulation}, most of the simulations in this paper have a set-up corresponding to the
experiment of Eldrup, Zinkle and Singh \textit{et al.}
\cite{zinkle2006microstructure,eldrup2002dose}. Accordingly, we use as reference conditions a dose rate of
$7\cdot10^{-7}$ dpa/s (except in Sec \ref{sec:flux}) and reach 0.73 dpa, we introduce 100 appm of spherical traps for vacancies, corresponding to
the amount of carbon and nitrogen in the iron used in the aforementioned experiment (except in Sec. \ref{sec:C_density}), and we set the temperature to 343 K in all cases. Unless otherwise stated, we use a simulation box size of $350\times400\times450\times a_0^3$.

\begin{figure}
 \centering
 \includegraphics[width=\columnwidth]{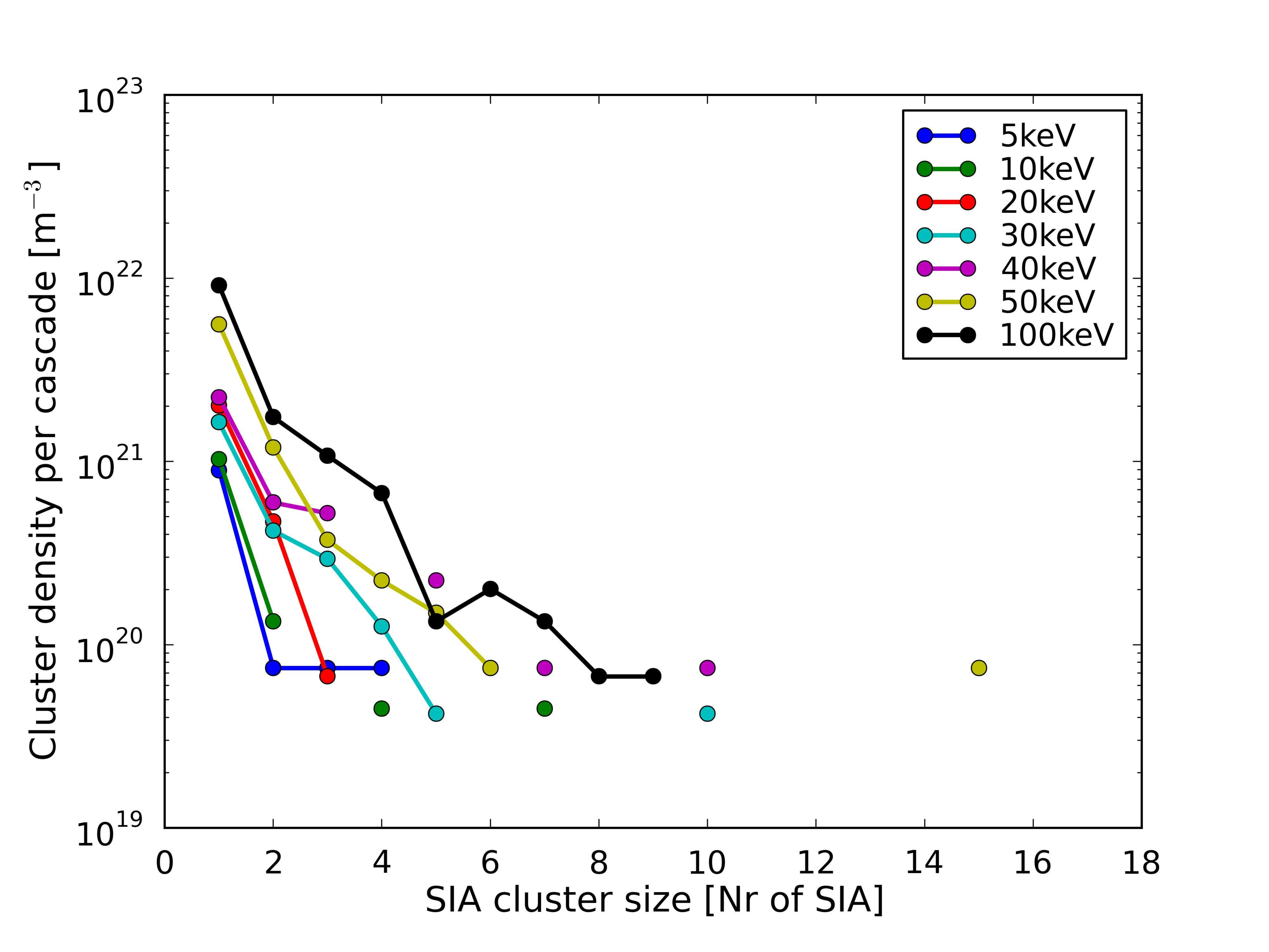}
 \caption{SIA cluster mean size distribution per cascade for different cascade energies in the used database \cite{stoller1996point,stoller1997primary,stoller2000statistical,
stoller2000evaluation,stoller2004secondary,stoller2000role}.}
 \label{cascade_sia_size_distribution.pdf}
\end{figure}

\begin{figure}
 \centering
 \includegraphics[width=\columnwidth]{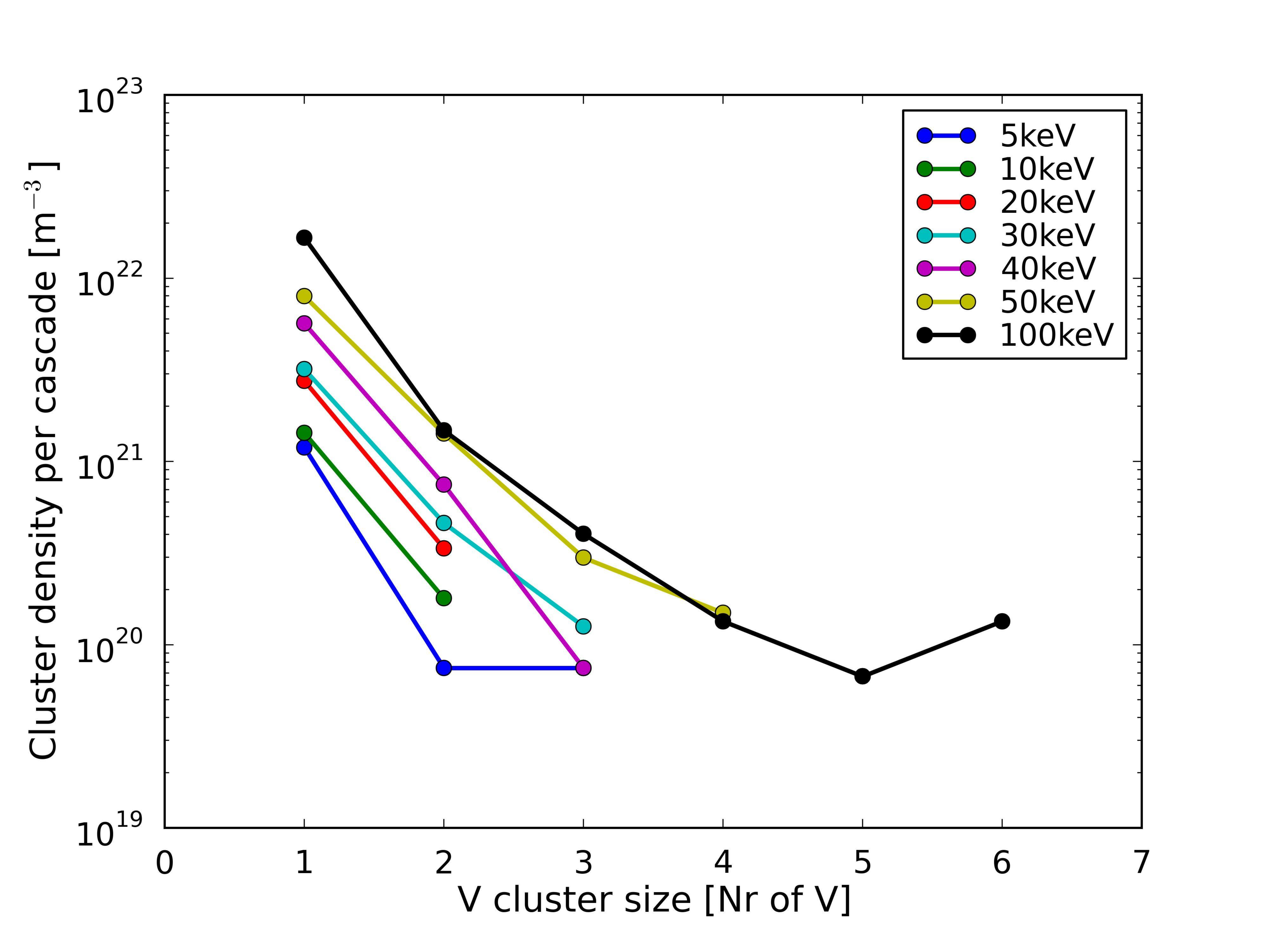}
 \caption{Vacancy cluster mean size distribution per cascade for different cascade energies in the used database \cite{stoller1996point,stoller1997primary,stoller2000statistical,
stoller2000evaluation,stoller2004secondary,stoller2000role}.}
 \label{cascade_vac_size_distribution.pdf}
\end{figure}

For simulating neutron irradiation, debris of vacancy and SIA objects of different sizes are randomly chosen from a database of displacement cascades \cite{stoller1996point,stoller1997primary,stoller2000statistical,
stoller2000evaluation,stoller2004secondary,stoller2000role} and randomly introduced into the system at a certain rate per time and volume. The displacement cascades were simulated using the Finnis-Sinclair potential \cite{finnis1984simple} and the considered cascade energies are 5 keV, 10 keV, 20 keV, 30 keV, 40 keV, 50 keV and 100 keV. The cluster size distribution per cascade, as used in our simulations, are shown for the SIA clusters in Fig. \ref{cascade_sia_size_distribution.pdf} and for the vacancy clusters in Fig. \ref{cascade_vac_size_distribution.pdf}. Vacancy clusters and SIA clusters smaller than size 150 have spherical shapes defined by a capture radius that depends on the cluster type and size (See \cite{jansson2013simulation} for details). The capture radius for a single vacancy is 4.315 Å and for a single SIA 6.396 Å. If two clusters overlap, they will recombine. The accumulated dpa is calculated using the NRT formula \cite{domain2004kinetic,norgett1975proposed}:
\begin{equation}
 dpa = \frac{0.8 E_{MD}}{2E_D},
\end{equation}
where $E_{MD}$ is the cascade energy and $E_D = 40$ eV is the displacement threshold energy for Fe. 

Vacancy and SIA clusters have a mobility that depends on their cluster size, as described in detail in \cite{jansson2013simulation}. Vacancy clusters migrate randomly in 3D. SIA clusters change direction according to a rotation energy, derived in terms of a Boltzmann expression. SIA clusters of size 1 have a rotation energy $\eta=0$ and will change direction at every jump, like the vacancy clusters. Larger SIA clusters will have a rotation energy that increases gradually with size and their migration direction will change less frequently. SIA clusters of size 12 and above have a rotation energy of 1 eV and will migrate fully in 1D along their Burgers vectors, at the simulation temperature considered in this work. If two SIA clusters meet, the resulting cluster will assume the Burgers vector and migration direction of the larger parent cluster. The rotational energy of the new cluster will depend on the cluster size. Only $\langle 111 \rangle$ SIA clusters are considered.

Spherical sinks with a sink strength corresponding to a dislocation
density of $10^{12}$ m$^{-2}$ are introduced as well. Defects are removed from the system when the
capture radii of the defect and the sink overlap. The effect of grain boundaries
is allowed for using the algorithm described in \cite{malerba2007object}.

It is known that C forms complexes with vacancies and that these complexes, in turn, trap SIA clusters \cite{anento2013carbon}. These effects are introduced in the
system using generic spherical traps for SIA or vacancies. When the capture
radii of a defect and a trap overlap, the two will be bound together with a
specified trapping energy $E^\delta_t$, which depends on whether the defect is of vacancy
($\delta=v$) or  SIA ($\delta=i$) type and on the size of the defect, as seen in Table
\ref{SIA_trapping_energies}. Above a threshold size $N_{th}$, we use a strong binding
energy for SIA clusters, $E^i_t=1.2$ eV, that can be associated with the binding energy of a CV$_2$ complex bound to the centre of a SIA cluster, and comparable to MD results \cite{anento2013carbon}. Below $N_{th}$, we use the binding energy with a single C atom. This choice allows for the fact that the vacancies in CV$_2$ complexes will only recombine if interacting with the edge of an SIA cluster and this type of interaction is the more likely, the smaller the cluster. 
\begin{table}
\centering
\caption{Trapping energy for the SIA traps at $T=343 K$. The values
for SIA clusters of sizes 1--4
and the values for vacancy cluster of sizes 1--6 are from DFT calculations
\cite{becquart2011p60}.}
\label{SIA_trapping_energies}
\begin{tabular*}{\columnwidth}{@{\extracolsep{\fill}} l c c}
\toprule
$N^i$ 	& SIA $E^i_t$ [eV]		& Vac. $E^v_t$ [eV]\\
\midrule
1 	& 0.17		 		& 0.65\\
2 	& 0.28		 		& 1.01\\
3 	& 0.36		 		& 0.93\\
4 	& 0.34		 		& 0.96\\
5	& 0.60		 		& 1.23\\
6	& 0.60				& 1.20\\
7--$N_{th}$	& 0.60			& 0.00\\
$N_{th}<$& 1.20		 		& 0.00\\
\bottomrule
\end{tabular*}
\end{table}

\section{Results}\label{sec:results}

\subsection{The effect of traps}\label{sec:traps}

\begin{figure}
 \centering
 \includegraphics[width=\columnwidth]{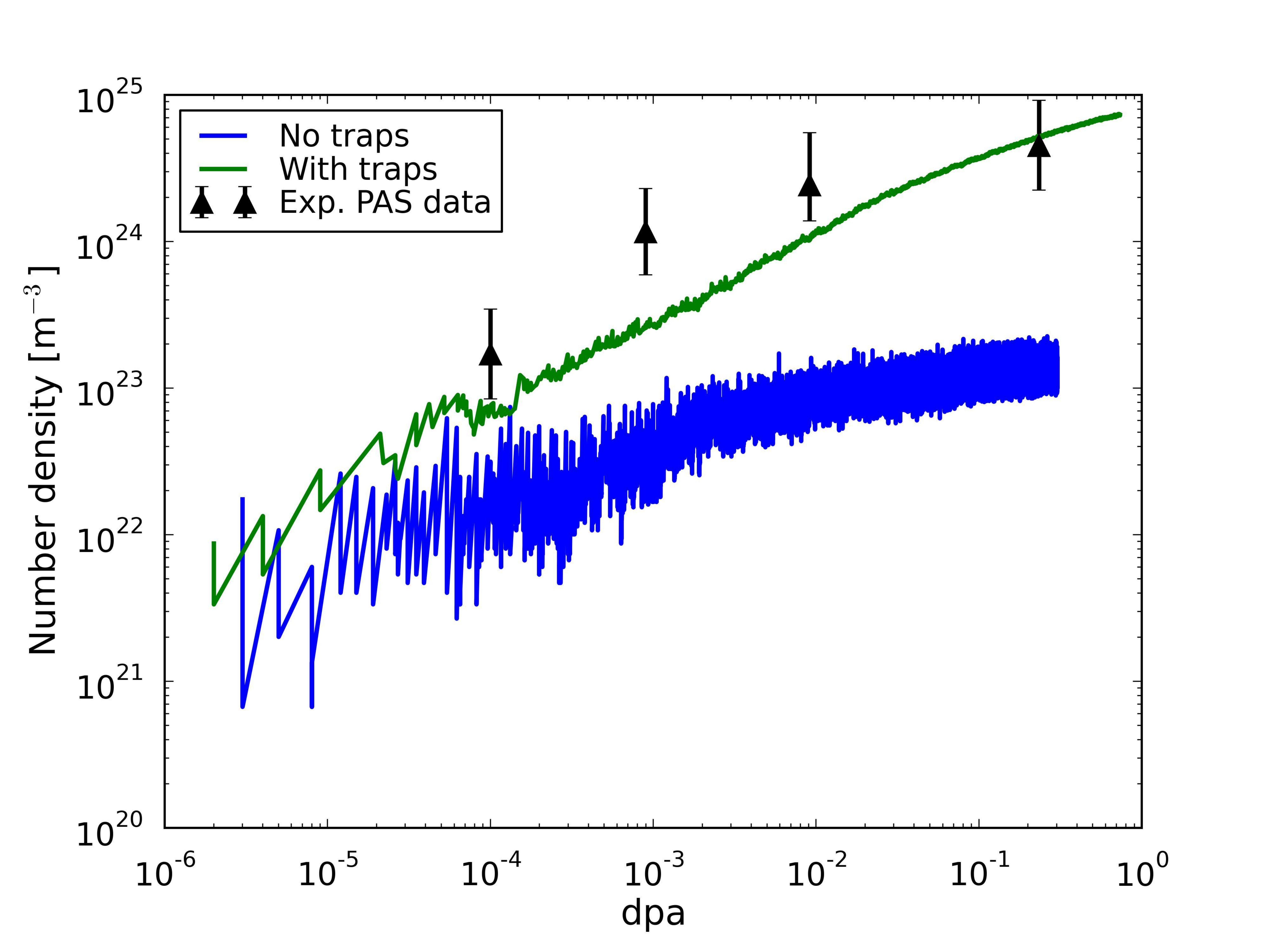}
 \caption{The vacancy cluster number density versus dpa with and without traps for vacancies and SIA clusters. The  experimental PAS data are from \cite{eldrup2002dose}.}
 \label{R20131230_vac.pdf}
\end{figure}

\begin{figure}
 \centering
 \includegraphics[width=\columnwidth]{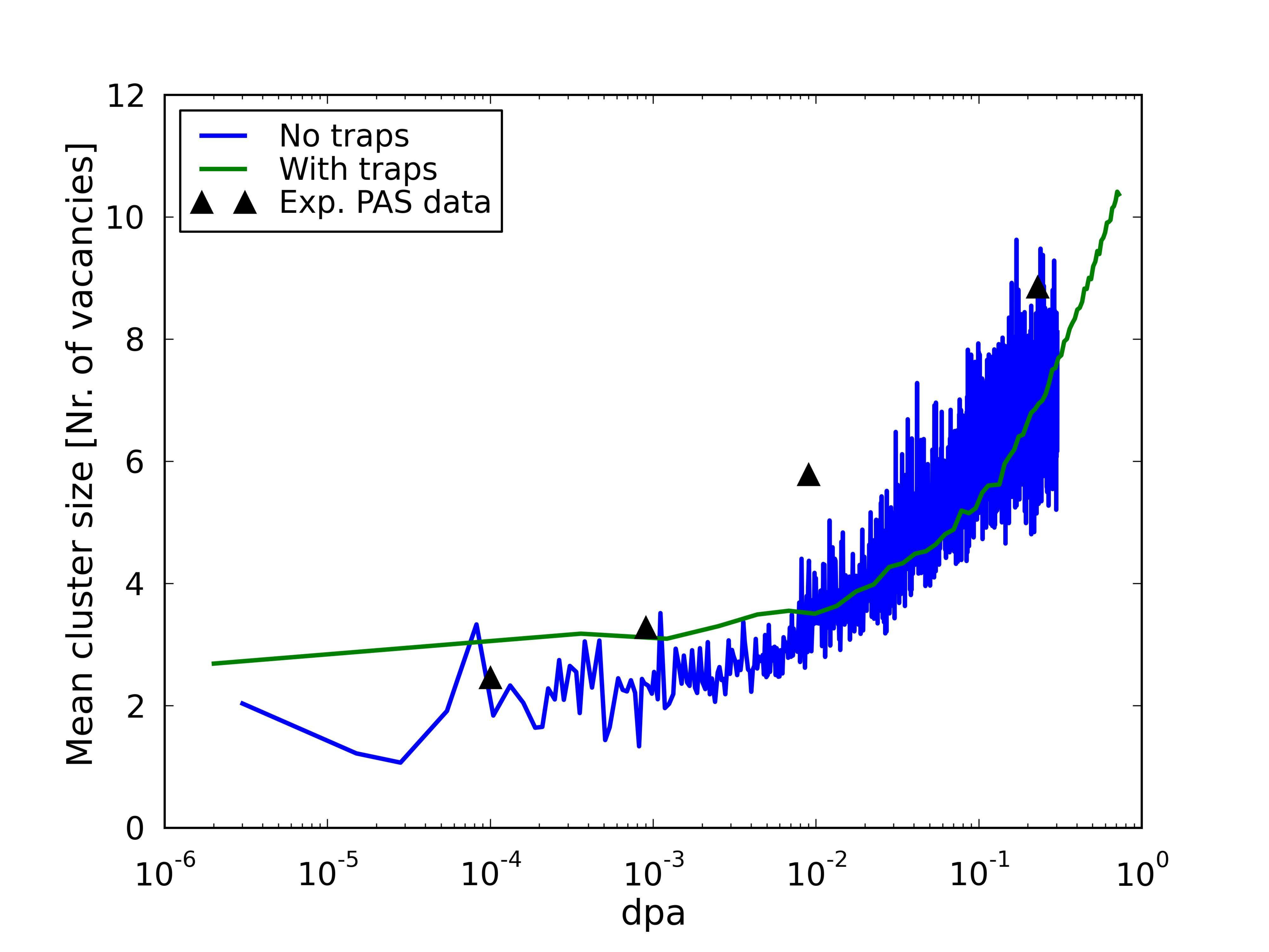}
 \caption{The mean vacancy cluster size versus dpa. The  experimental PAS data are from \cite{eldrup2002dose}.}
 \label{R20131230_vac_mean_size_evolution.pdf}
\end{figure}

Traps for vacancy and SIA clusters were shown in \cite{jansson2013simulation} to play a key role in the nanostructure evolution, where simulations of irradiation of Fe-C at 343 K obtained good agreement with experiments \cite{zinkle2006microstructure,eldrup2002dose}. In this study the traps were removed, which results in significantly lower vacancy cluster density, as shown in Fig. \ref{R20131230_vac.pdf}, even though the vacancy cluster mean size evolution is not significantly affected, as shown in Fig. \ref{R20131230_vac_mean_size_evolution.pdf}. Without traps, no visible SIA clusters appear, contrary to the experimental data \cite{eldrup2002dose} and the results in \cite{jansson2013simulation}, where traps were used. 

\subsection{The effect of carbon content}\label{sec:C_density}

We studied the model's sensitivity to C content by varying the concentration of SIA and
vacancy traps from 1 appm to 200 appm, 100 appm corresponds to the reference concentration \cite{zinkle2006microstructure,eldrup2002dose}, which the model was fine-tuned to reproduce. It is unlikely that the C content in the matrix would exceed these concentrations, even in steels. The SIA and vacancy trapping energies have a size dependency as described in Table \ref{SIA_trapping_energies} and the threshold $N_{th}=29$.

\begin{figure}
 \centering
  \includegraphics[width=\columnwidth]{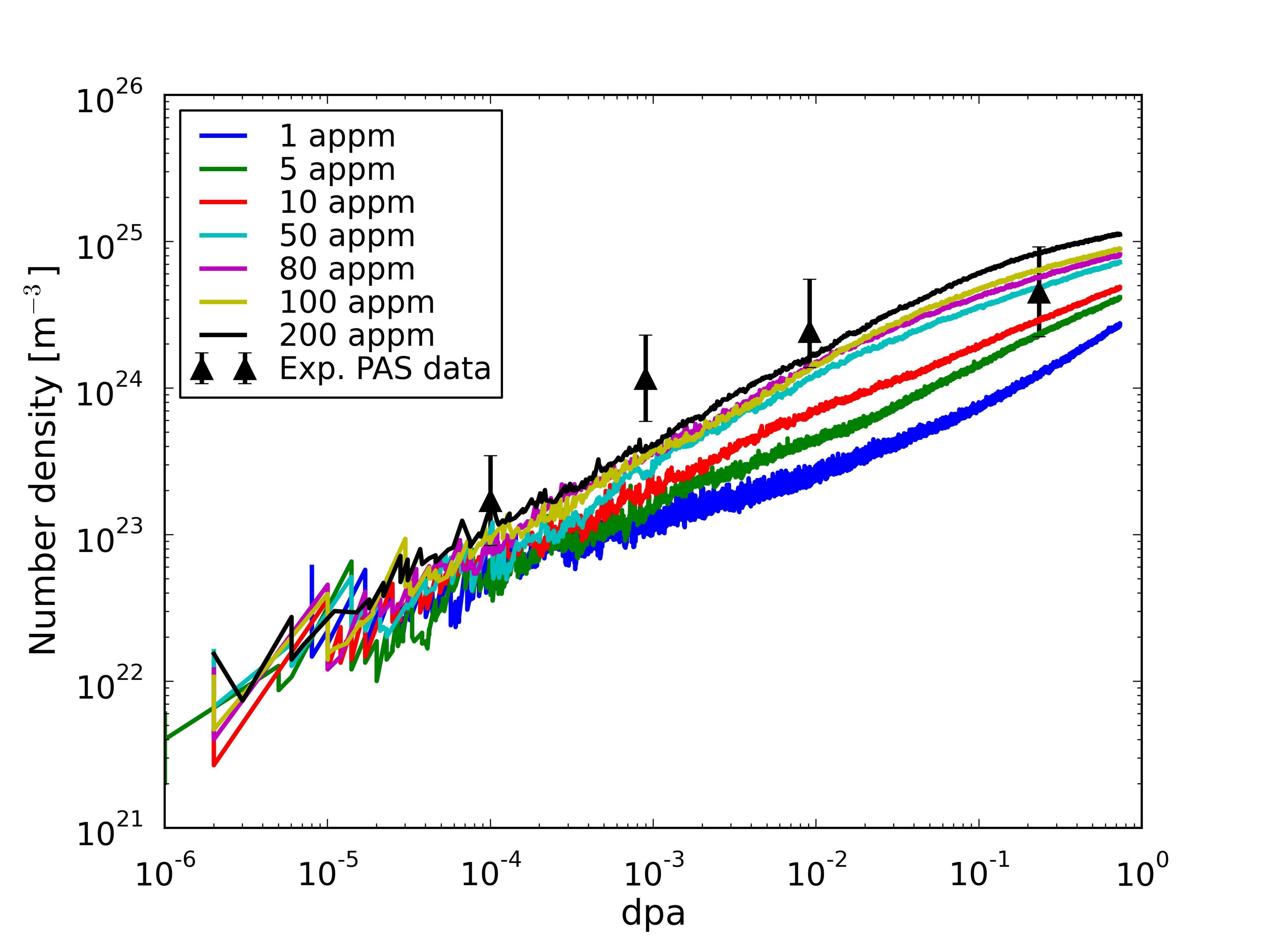}
  \caption{Sensitivity of the results to trap concentration: the vacancy cluster density versus dpa. The case with 100 appm traps corresponds to the carbon content in the reference experiment \cite{eldrup2002dose}.}
  \label{R20120905-0_vac_dens.pdf}
\end{figure}

\begin{figure}
 \centering
  \includegraphics[width=\columnwidth]{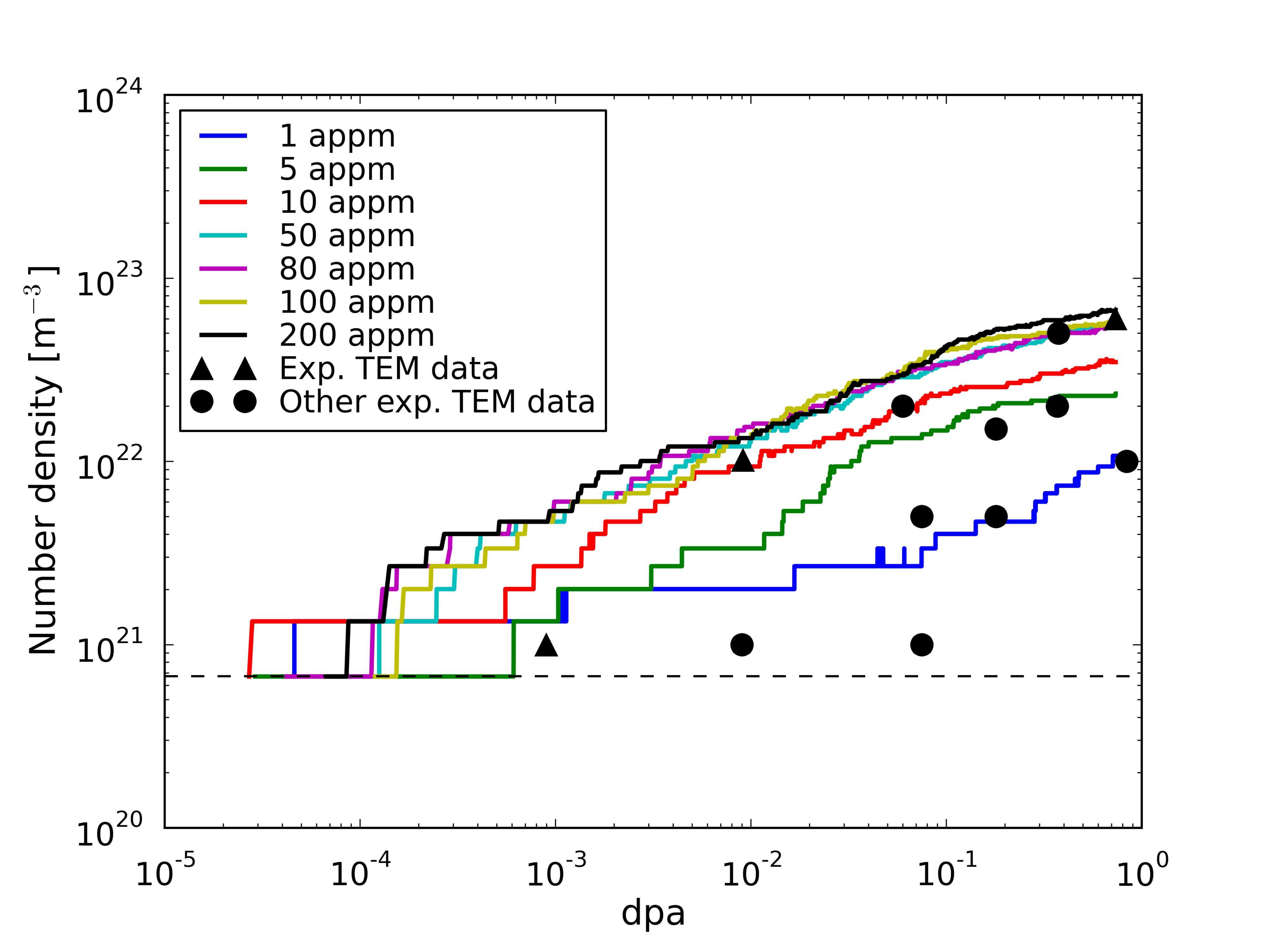}
  \caption{Sensitivity of the results to trap concentration: the visible SIA density versus
dpa. The reference experimental data are denoted with triangles
  \cite{zinkle2006microstructure}. Included in the graph are also data from
other
comparable irradiation experiments in Fe-C (bullets)
\cite{singh1999effects,eyre1965electron,bryner1966study,robertson1982low,
horton1982tem,takeyama1981}. See \cite{malerba2011review} for full
details. The dotted line gives the density for one visible cluster in the simulation
box.}
  \label{R20120905-0_visible_SIA.pdf}
\end{figure}

In Fig. \ref{R20120905-0_vac_dens.pdf}, the vacancy cluster density evolution for the
different trap concentrations is shown. It is observed that the higher the trap concentration, the higher the density of clusters, although this effect is strong for low trap contents and tends to saturate for higher contents. This saturation is even more clear for the visible SIA cluster density evolution shown in Fig, \ref{R20120905-0_visible_SIA.pdf}: it appears that the correct trend is reached already with only 10 appm traps and higher trap densities gives no higher SIA density, although there is a shift to higher dose for the appearance of the first visible clusters. It should be noted, however, that the fact that other experimental points than those used here as reference lie on curves corresponding to lower C content should not be ascribed to less C in those experiments. Instead, it is probably the consequence of the fact that the resolution of microscopes has been increasing over the years \cite{malerba2011review}.

\begin{figure}
 \centering
  \includegraphics[width=\columnwidth]{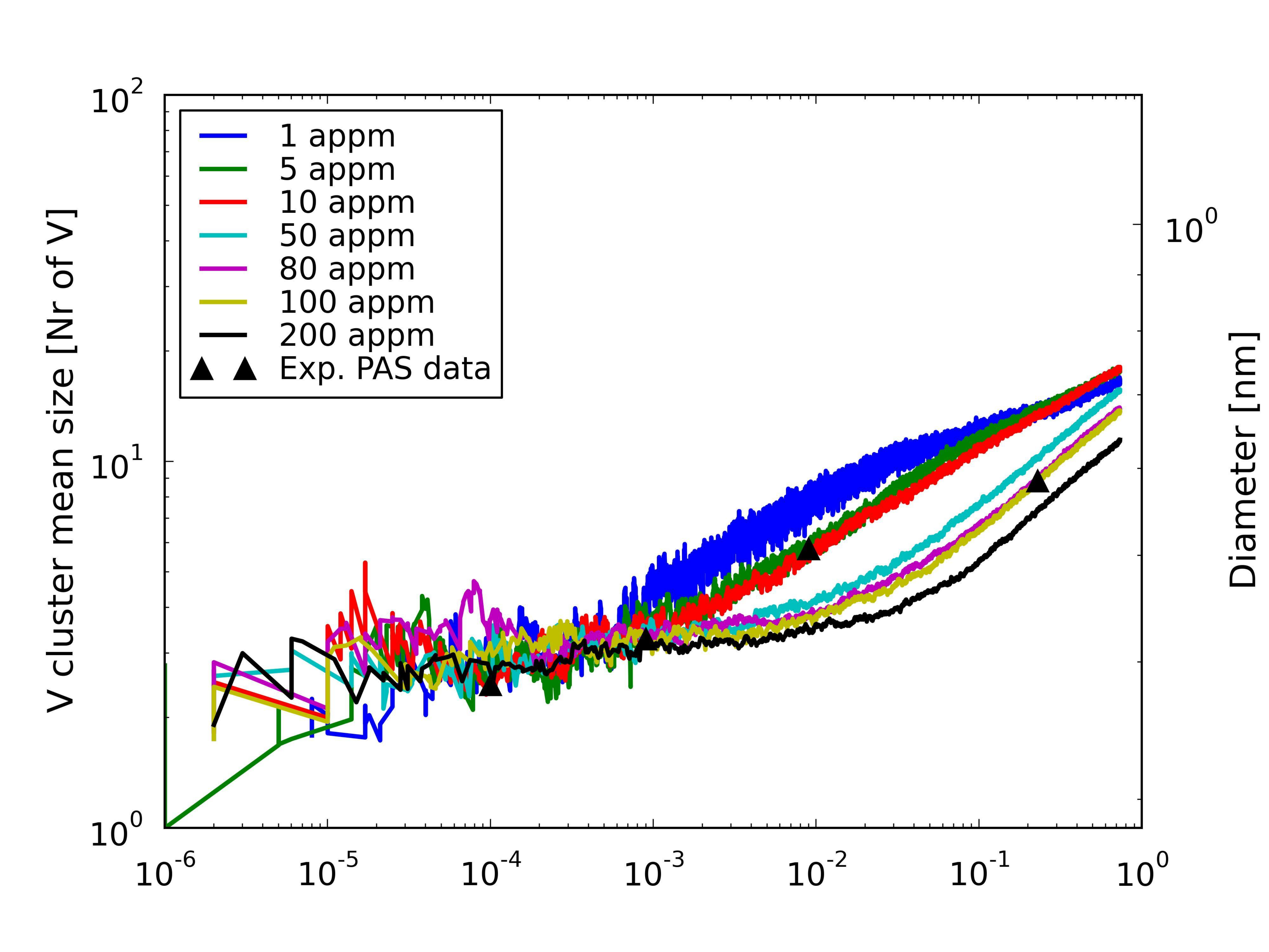}
    \caption{Sensitivity of the results to trap concentration: vacancy cluster mean size versus dpa. The reference experimental data are for $\sim$100 appm C \cite{eldrup2002dose}.}
  \label{R20130129-c_321841_vac_mean_size_evolution.pdf}
\end{figure}

\begin{figure}
 \centering
  \includegraphics[width=\columnwidth]{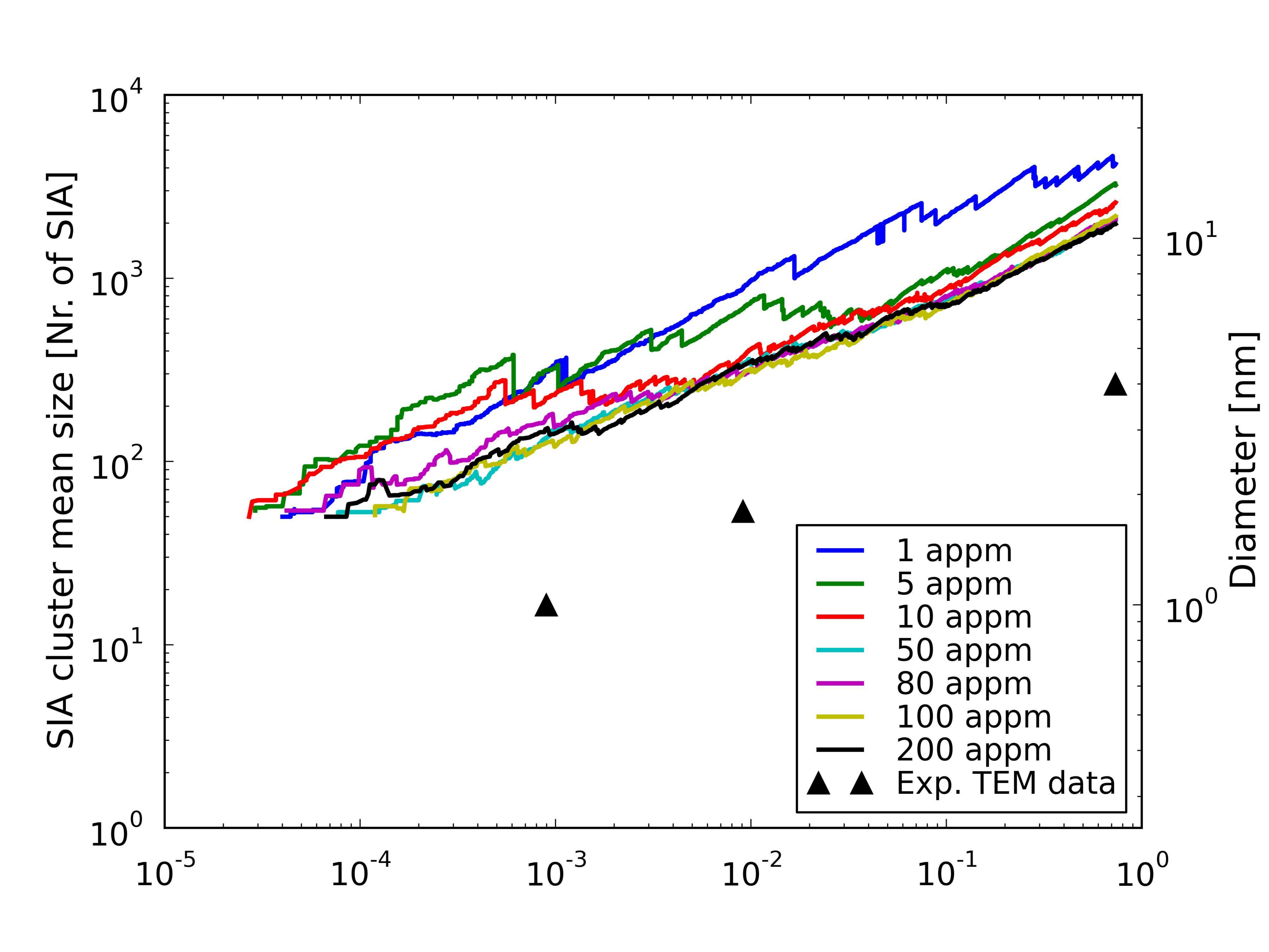}
    \caption{Sensitivity of the results to trap concentration: SIA cluster mean size versus dpa. The reference experimental data are for $\sim$100 appm C \cite{zinkle2006microstructure}.}
  \label{R20130129-c_sia_mean_size_evolution.pdf}
\end{figure}

In Fig. \ref{R20130129-c_321841_vac_mean_size_evolution.pdf} and \ref{R20130129-c_sia_mean_size_evolution.pdf} the cluster mean size evolution versus dpa for vacancy and SIA clusters, respectively, are shown. The general trend for both kinds of clusters is that the mean size increase with decreased C concentration, even though the effect is very small for vacancy clusters. The 
vacancy mean sizes do not differ more than 0.2 nm at most from the experimental data \cite{eldrup2002dose} for any trap concentration. The main effect of reducing the trap concentrations is that the negative curvature in the size evolution is anticipated to lower dose. For 100 appm traps, the marginal is even less than that and thus in good agreement with the experiment. For the SIA cluster sizes, it can be observed that the sizes do not change much for C concentrations above 5 appm. Above 80 appm, the mean size evolutions are almost identical. The model overestimates the SIA cluster mean size, about a factor of 5 higher than the experimental TEM data, but, on the other hand, the experimental data give surprisingly small mean sizes for the SIA clusters. Zinkle and Singh report a TEM resolution of $\sim$0.5 nm \cite{zinkle2006microstructure}, wheras 1.5 nm is a more common values and also the one we have choosen to use as our limit. Using a lower limit does not improve the results.

\subsection{The effect of the threshold} \label{sec:threshold}
  
The threshold, $N_{th}$, makes the energy with which SIA clusters are trapped
depend on the size of the trapped cluster. SIA clusters of size $5\leq N^i\leq
N_{th}$ are trapped with $E^i_t=0.6$ eV, which is the binding energy
between a C atom and an SIA cluster \cite{terentyev2011interaction}. For $N^i > N_{th}$ they are trapped with 1.2 eV, which corresponds to the strong binding energy between a CV$_2$ complex and the centre of a SIA cluster. This is summarized in Table \ref{SIA_trapping_energies}. $N_{th}$ is the only parameter of the model used for fine-tuning to fit the experimental data. It is therefore important to verify up to what extent the results depend on it.
  
\begin{figure}
  \centering
  \includegraphics[width=\columnwidth]{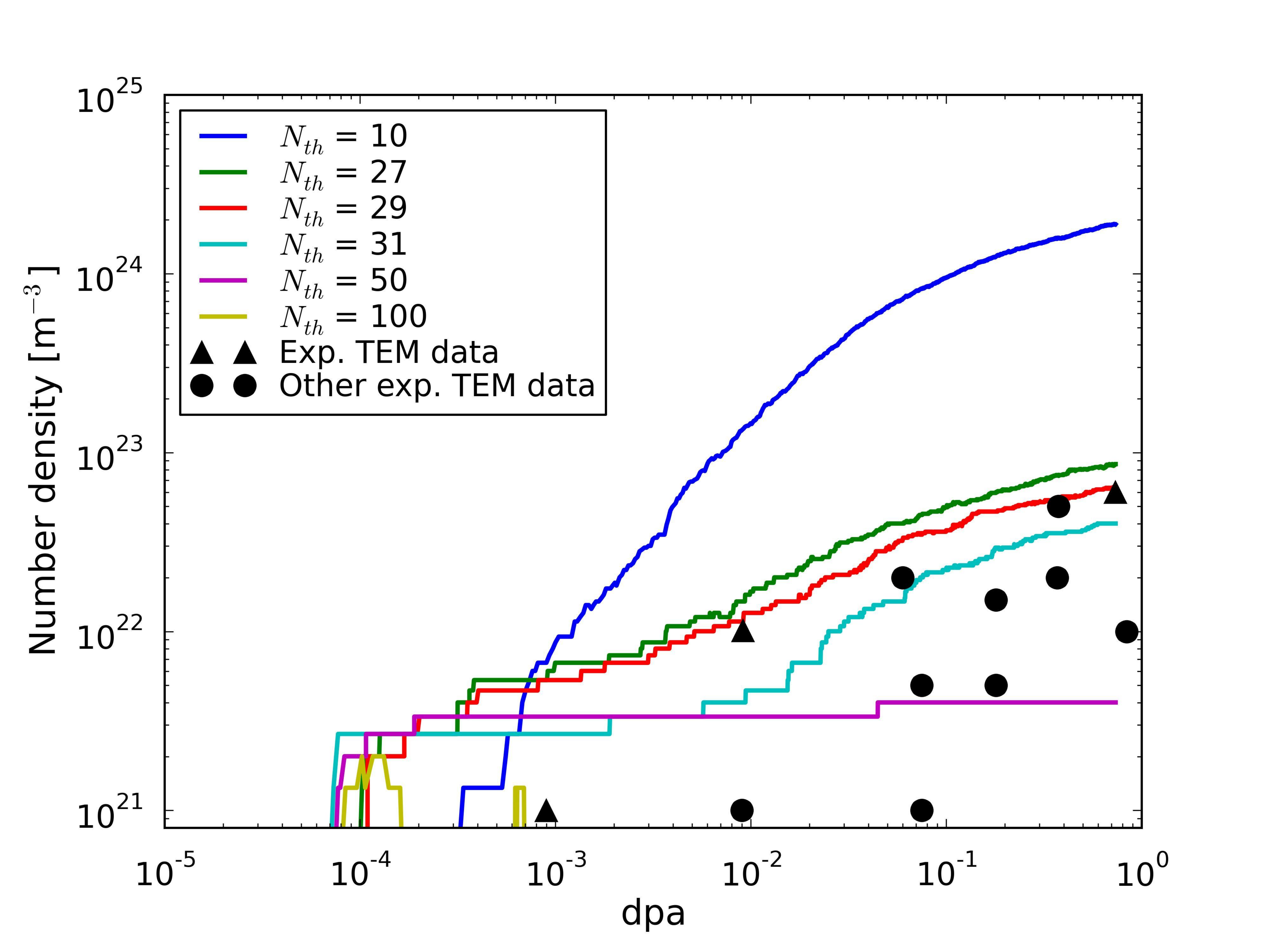}
  \caption{Number density of visible SIA versus dpa for different values of the
threshold parameter, $N_{th}$. The reference experimental data are denoted with
triangles
  \cite{zinkle2006microstructure}. Included in the graph are also data from
other
  comparable irradiation experiments in Fe-C (bullets)
  \cite{singh1999effects,bryner1966study,robertson1982low,
  horton1982tem,takeyama1981,eyre1965electron}. See \cite{malerba2011review} for full details.}
  \label{R20120423_visible_SIA_av.pdf}
  \end{figure}

\begin{figure}
 \centering
  \includegraphics[width=\columnwidth]{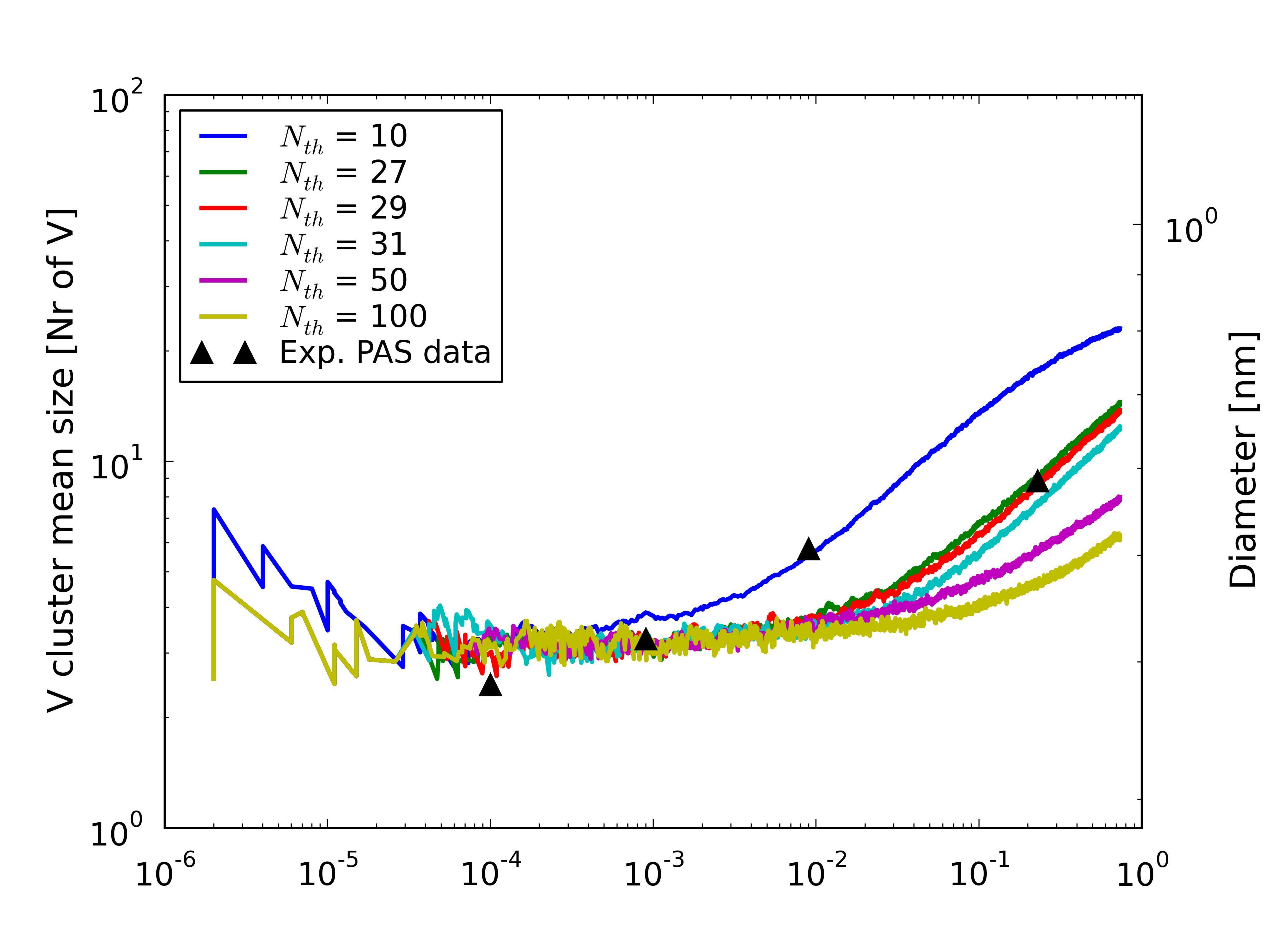}
    \caption{Sensitivity of the results to the threshold parameter, $N_{th}$: vacancy cluster mean size versus dpa. The experimental data are
from \cite{eldrup2002dose}.}
  \label{R20130107-a_vac_mean_size_evolution.pdf}
\end{figure}

\begin{figure}
 \centering
  \includegraphics[width=\columnwidth]{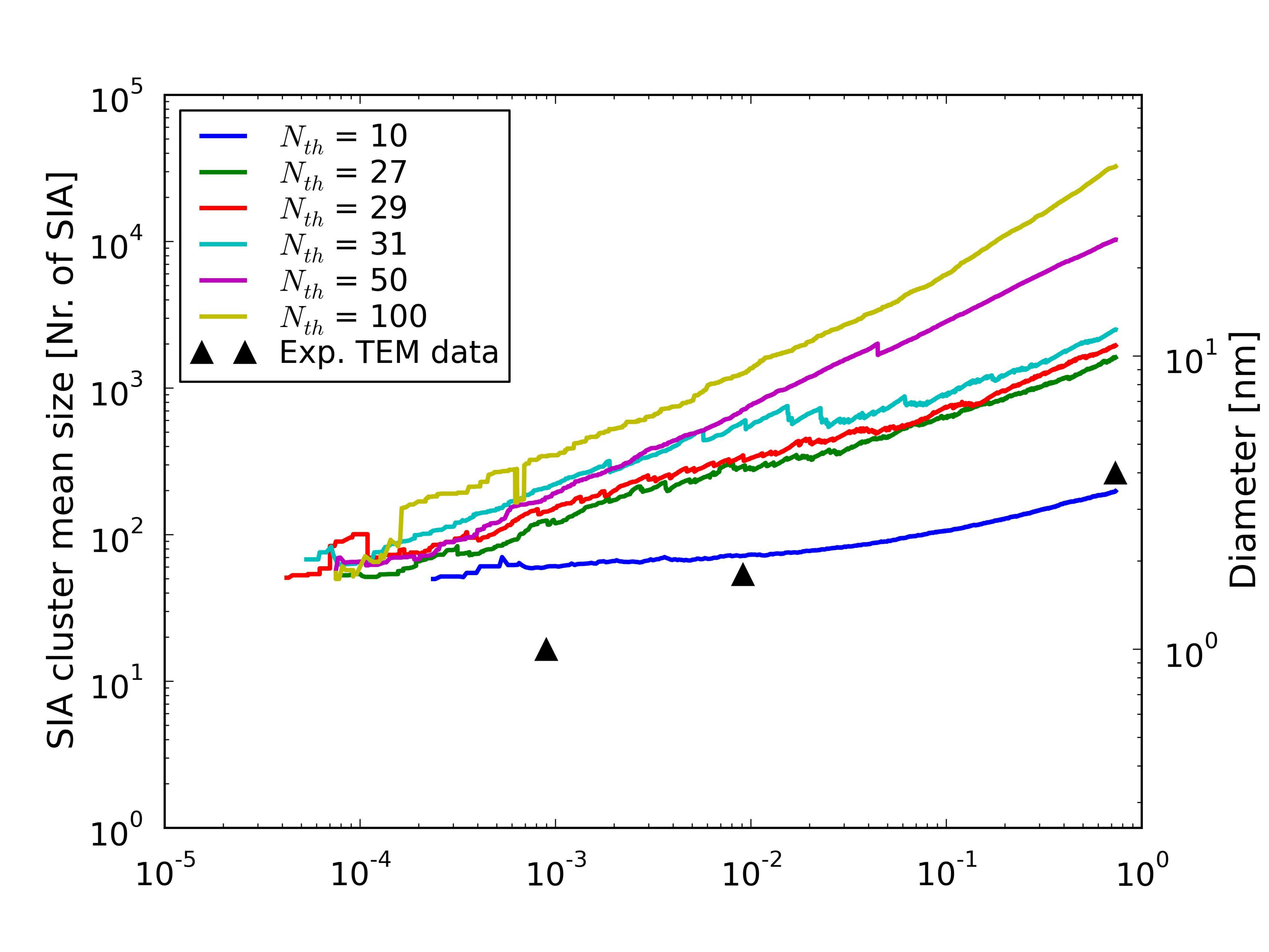}
    \caption{Sensitivity of the results to the threshold parameter, $N_{th}$: SIA cluster mean size versus dpa. The experimental data are
from \cite{zinkle2006microstructure}.}
  \label{R20130107-a_sia_mean_size_evolution.pdf}
\end{figure}

We used the simulation set-up corresponding to \cite{zinkle2006microstructure} with 100
appm traps for SIA and vacancy clusters, respectively.  
  In Fig. \ref{R20120423_visible_SIA_av.pdf}, the scaling with dose of the density of
visible SIA clusters is shown for different values of $N_{th}$. For comparison, the experimental data from \cite{singh1999effects,bryner1966study,robertson1982low,
  horton1982tem,takeyama1981,eyre1965electron} are also included.
The best fit to match the reference experiment from \cite{zinkle2006microstructure} is given by $N_{th}=29$, as already reported in
\cite{jansson2013simulation}. Higher values give lower densities, which remain acceptable, as compared to other experiments, up to 50. Very high values provide negligible densities, as a consequence of the fact that most clusters are very mobile and only a few remain trapped stably enough to manage to grow by absorbing the smaller ones. On the other hand, low values of $N_{th}$ shoot the number density by having the opposite effect: too many clusters remain stably trapped and act as nuclei of clusters growing above the visibility threshold. The vacancy cluster density follows the same trend with $N_{th}$, likely because of enhanced recombination with more mobile SIA clusters, while the mean vacancy cluster sizes are larger with smaller $N_{th}$ (\textit{Cf.} \ref{R20130107-a_vac_mean_size_evolution.pdf}). The mean SIA cluster sizes are larger with smaller $N_{th}$ (\textit{Cf.} \ref{R20130107-a_sia_mean_size_evolution.pdf}).

\subsection{Effect of the trapping energy of large SIA sizes}\label{sec:etlarge}
 
We have also analyzed the sensitivity of the results to the trapping energy, $E^i_t$ for
SIA clusters above the threshold size $N_{th}=29$. In principle such energy is fixed by the type of traps (C-vacancy clusters) with which the SIA clusters interact and by where the interaction occurs (centre versus periphery). However, since the value for this interaction energy comes from atomistic simulations, it is worth verifying which values are acceptable. We thus considered different values
for the $E_t^i$ above the threshold, from 0.8 eV to 2.0 eV.
 
\begin{figure}
 \centering
  \includegraphics[width=\columnwidth]{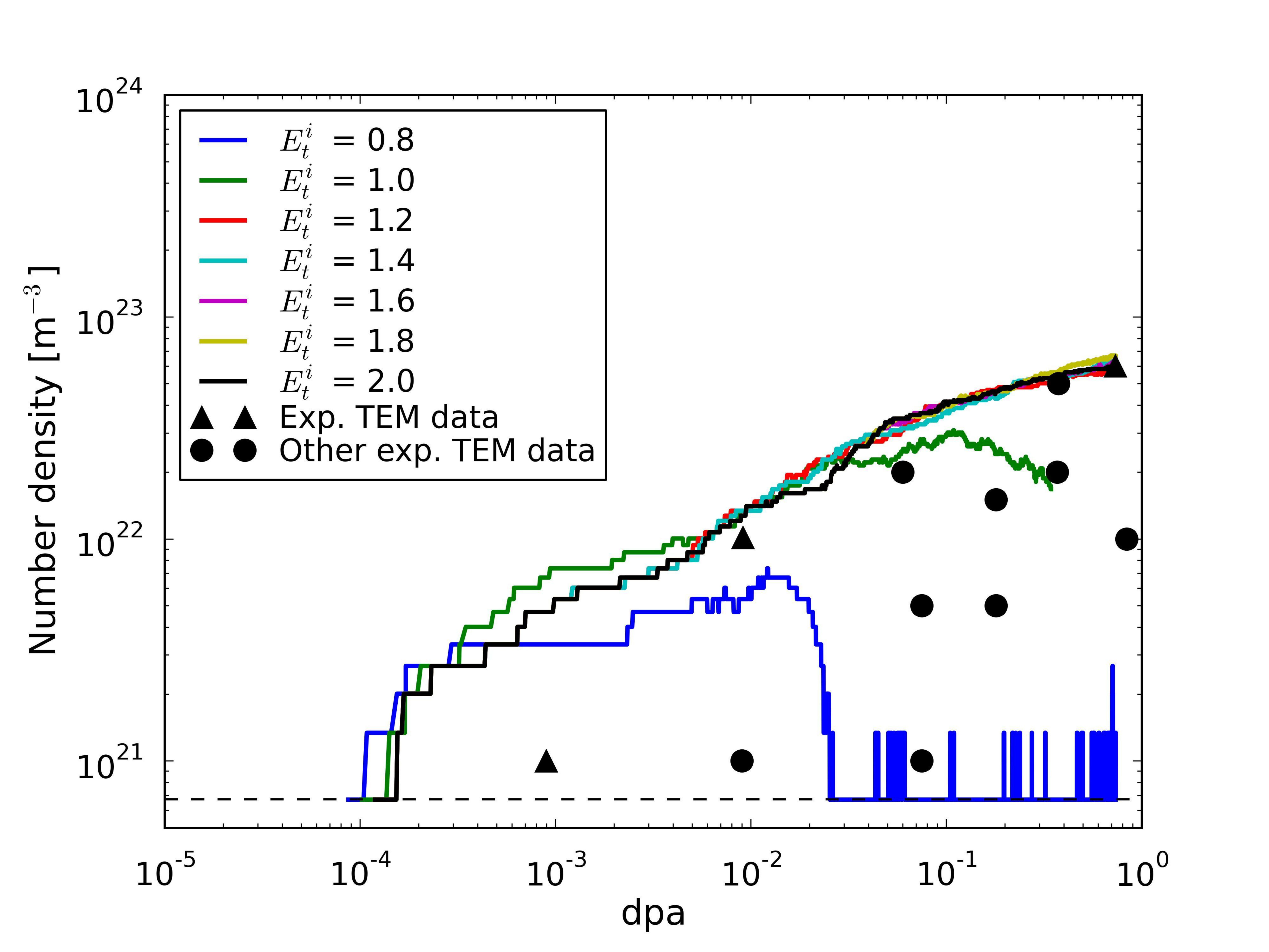}
  \caption{Sensitivity of the visible SIA number density to the value of the trapping energy above $N_{th}$. The reference experimental
data are denoted with triangles
  \cite{zinkle2006microstructure}. Included in the graph are also data from
other comparable irradiation experiments in Fe-C (bullets)
\cite{singh1999effects,bryner1966study,robertson1982low,
horton1982tem,takeyama1981,eyre1965electron}. See \cite{malerba2011review} for full
details. The dotted line gives the density for one visible cluster in the simulation
box.}
  \label{R20120905-e_visible_SIA.pdf}
\end{figure}

\begin{figure}
 \centering
  \includegraphics[width=\columnwidth]{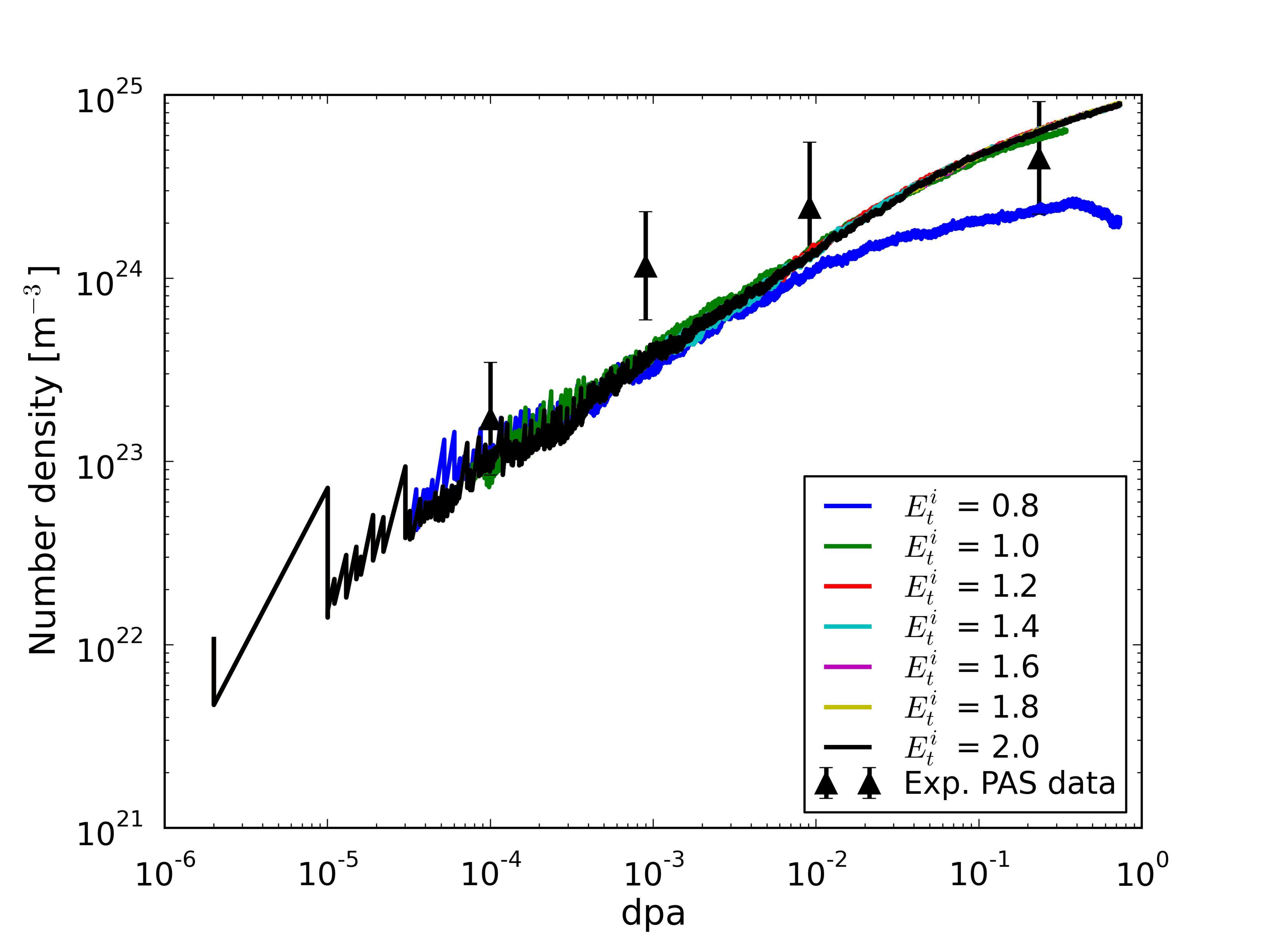}
  \caption{Sensitivity of the results to $E_t$ for large SIA clusters: the vacancy cluster density
versus dpa. The experimental data are from \cite{eldrup2002dose}.}
  \label{R20120905-e_vac.pdf}
\end{figure}

\begin{figure}
 \centering
  \includegraphics[width=\columnwidth]{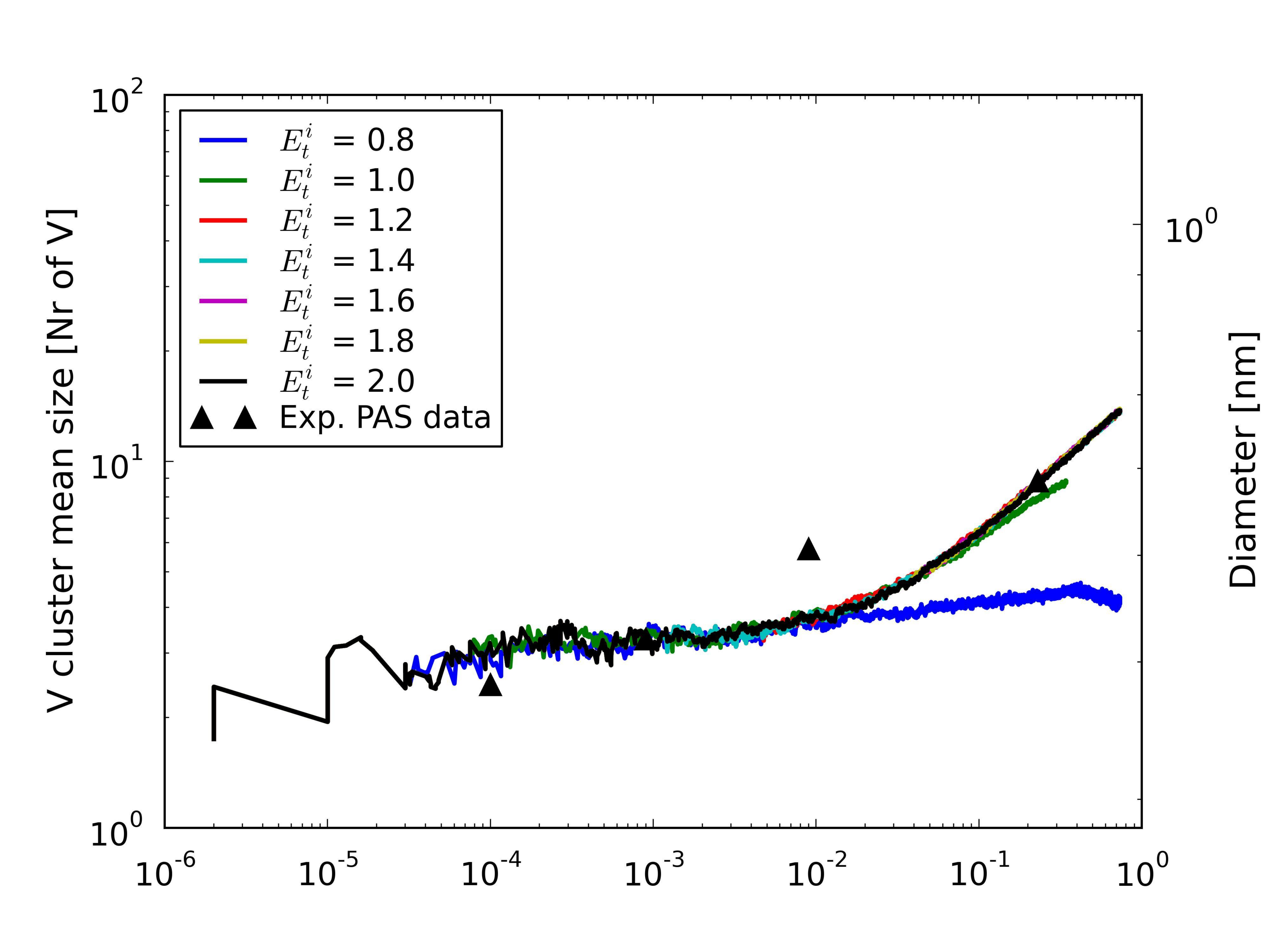}
    \caption{Sensitivity of the result to the trapping energy above $N_{th}$: vacancy cluster mean size versus dpa. The experimental data are
from \cite{eldrup2002dose}.}
  \label{R20130129-e_321841_vac_mean_size_evolution.pdf}
\end{figure}

\begin{figure}
 \centering
  \includegraphics[width=\columnwidth]{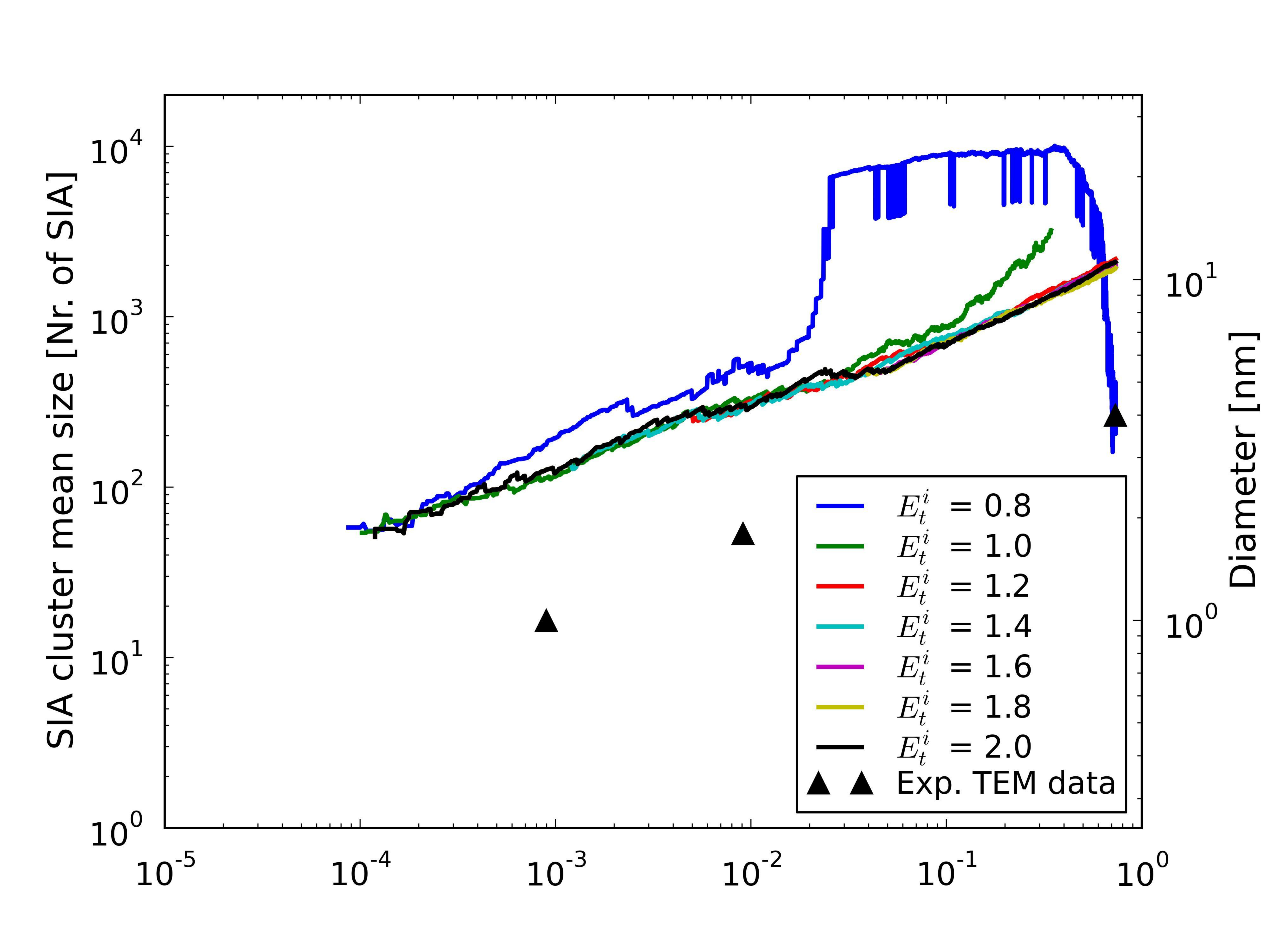}
    \caption{Sensitivity of the results to the trapping energy above $N_{th}$: SIA cluster mean size versus dpa. The experimental data are
from \cite{zinkle2006microstructure}.}
  \label{R20130129-e_sia_mean_size_evolution.pdf}
\end{figure}

We see that higher $E^i_t$ gives higher densities both for visible SIA (\textit{Cf.} Fig. \ref{R20120905-e_visible_SIA.pdf}) and for
vacancies (\textit{Cf.} Fig. \ref{R20120905-e_vac.pdf}). For visible SIA, however, we see no significant difference when $E^i_t$ exceeds 1.2 eV. The effect on vacancy cluster density is at that point totally negligible.
Considering the mean cluster sizes, the evolutions for both vacancy and SIA clusters (Fig. \ref{R20130129-e_321841_vac_mean_size_evolution.pdf} and \ref{R20130129-e_sia_mean_size_evolution.pdf}, respectively) are very similar for $E_t^\delta=$ 1.0--2.0 eV and show good agreement with the experimental data (\cite{eldrup2002dose} and \cite{zinkle2006microstructure}, respectively) in the case of vacancy clusters and fair agreement for the SIA clusters. With $E_t^\delta=0.8$ eV, the trend differs considerably from the experimental data for both kinds of defects. Essentially, the effect of lowering the trapping energy is that, after the initial buildup of cluster populations, above a certain dose ($\sim$0.01 dpa), massive recombination starts occurring, thereby reducing both size and density of clusters.

\subsection{Study of the effect of the dose rate}\label{sec:flux}

To conclude this study, we investigated the effect of an environmental, rather than model, parameter, namely the flux, by varying the dose rate from $1.0\cdot 10^{-11}$ to $1.0\cdot 10^{-3}$ dpa/s. The former value corresponds to a typical flux on the RPV walls in commercial nuclear power plants, whereas the latter value corresponds to ion irradiation. The importance of this parameter resides in the fact that as yet it is not established whether or not one should expect significant differences between irradiation in materials test reactors as opposed to surveillance specimens irradiated during operation. The box size was here $150\times200\times250\times a_0^3$ and the temperature was 343 K. Traps for SIA were introduced with a density of 100 appm and a trapping energy of 1.0 eV. The set-up is thus similar to the experimental set-up in \cite{eldrup2002dose,zinkle2006microstructure}. The simulation was stopped after 0.23 dpa.

\begin{figure}
 \begin{center}
  \includegraphics[width=\columnwidth]{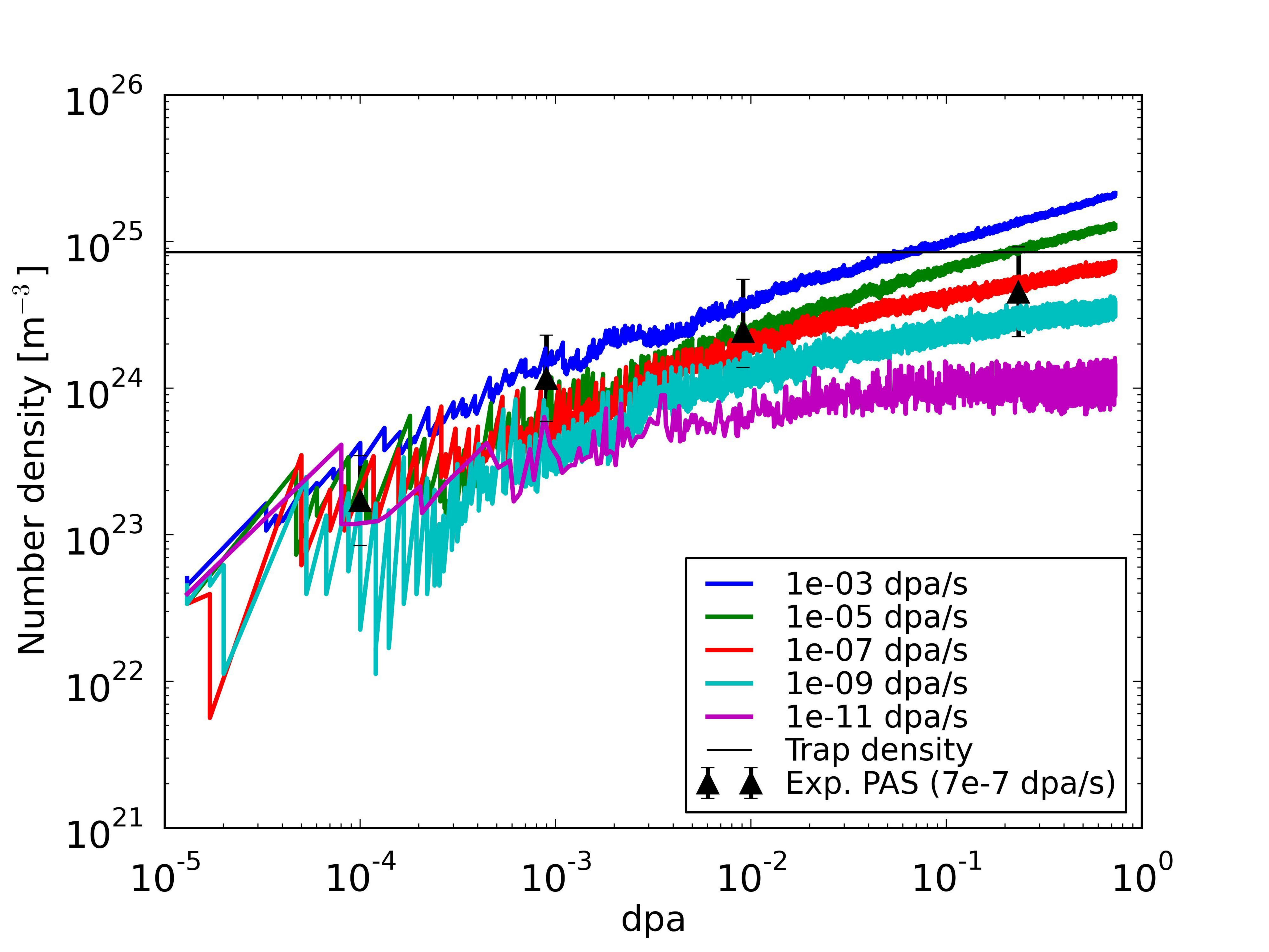}
  \caption{Effect of dose on the vacancy density evolution. The experimental
data are from \cite{eldrup2002dose} and correspond to a dose rate of $7\cdot10^{-7}$ dpa/s. The black line is the trap density.}
  \label{R20111202-1_vac_dens.pdf}
 \end{center}
\end{figure}

\begin{figure}
 \begin{center}
  \includegraphics[width=\columnwidth]{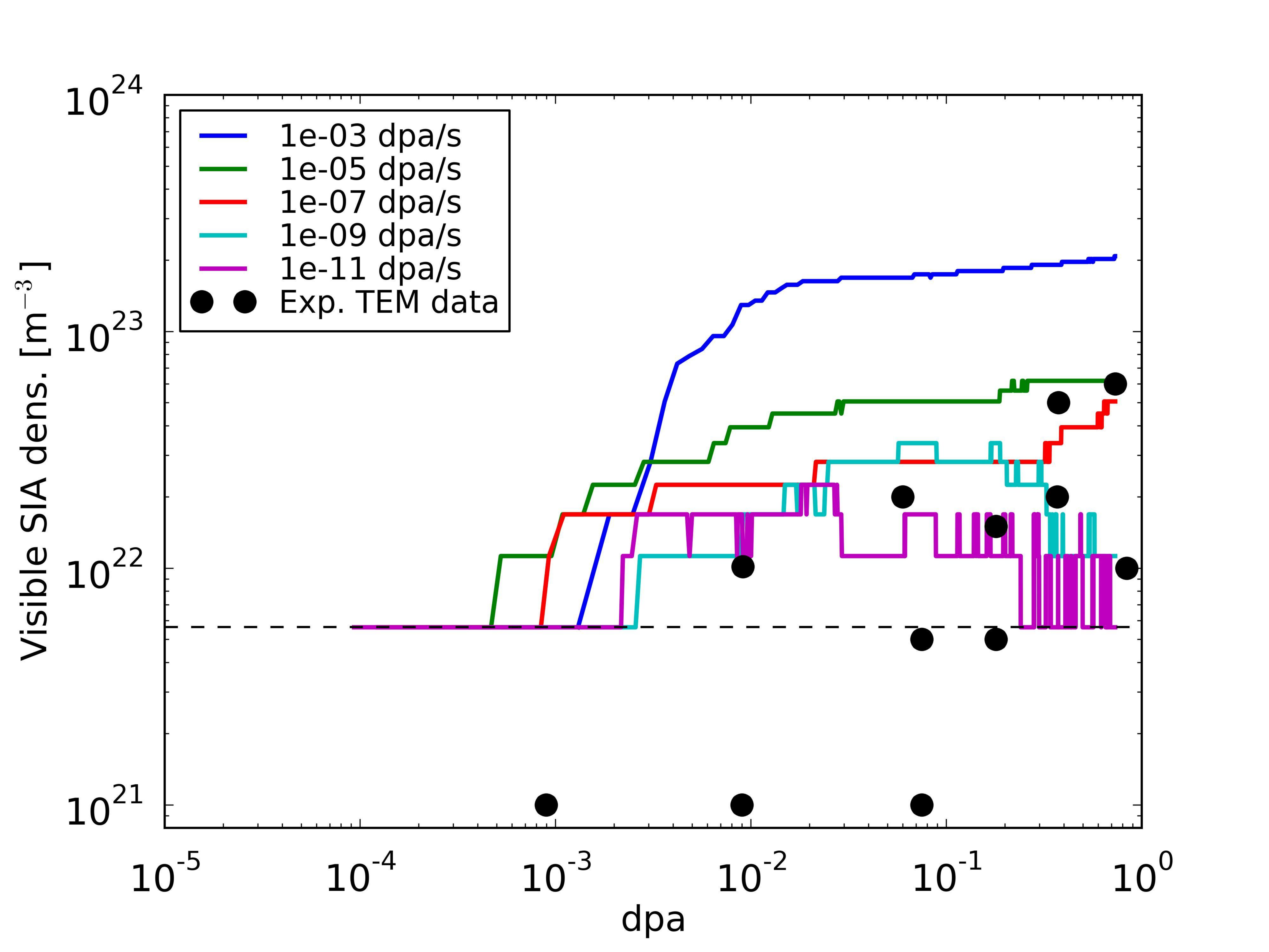}
  \caption{Effect of dose rate on the visible SIA cluster density evolution. The dotted line gives the density for one visible cluster in the simulation
box. The experimental data are from 
\cite{zinkle2006microstructure,singh1999effects,bryner1966study,
robertson1982low,horton1982tem,takeyama1981,eyre1965electron}. See \cite{malerba2011review} for
full details.}
  \label{R20120520-00_SIA_dens.pdf}
 \end{center}
\end{figure}

\begin{figure}
 \centering
  \includegraphics[width=\columnwidth]{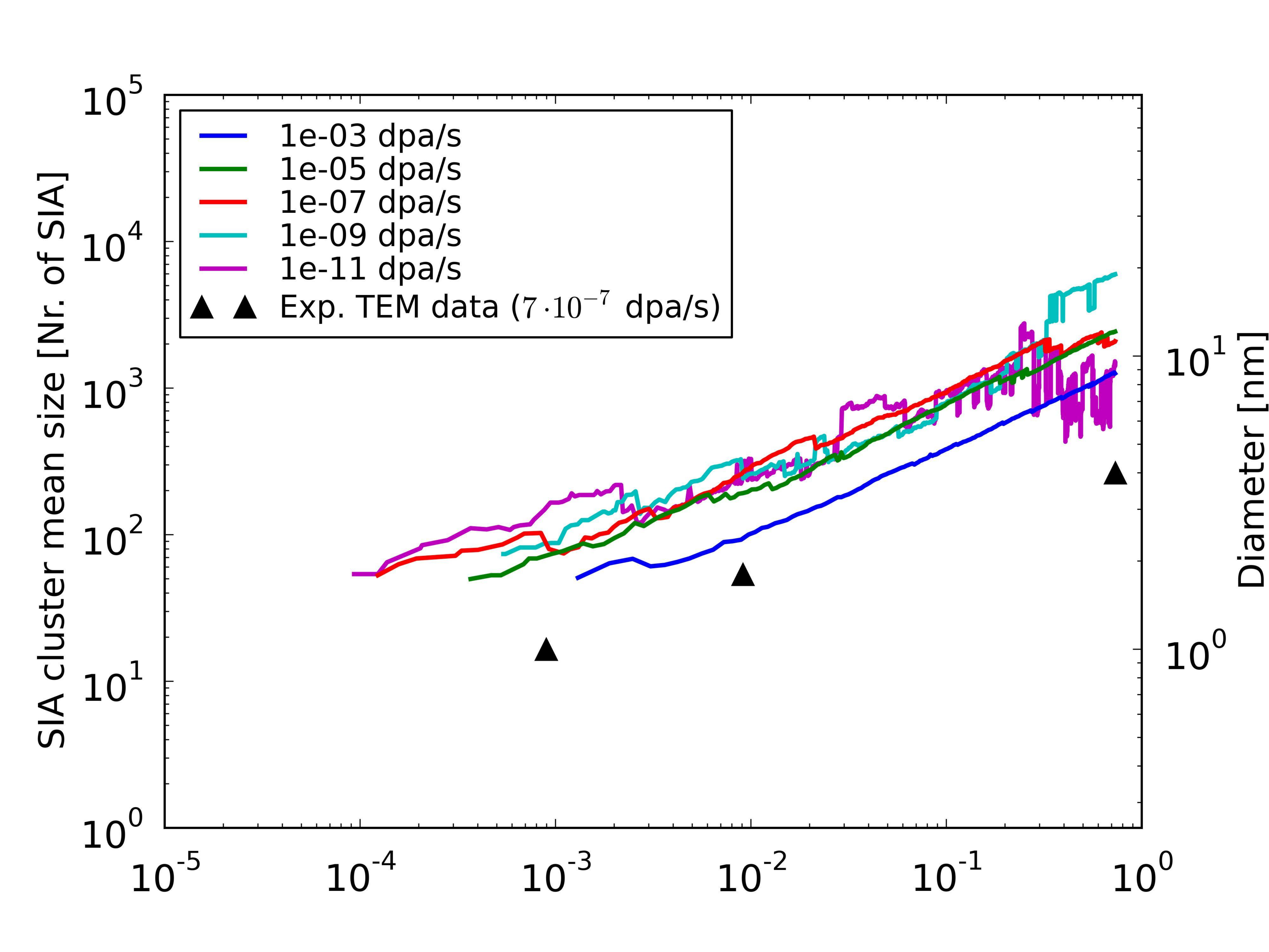}
    \caption{Sensitivity of the results to the dose rate: SIA cluster mean size versus dpa. The experimental data are
from \cite{zinkle2006microstructure}.}
  \label{R20130129-a1_sia_mean_size_evolution.pdf}
\end{figure}

\begin{figure}
 \centering
  \includegraphics[width=\columnwidth]{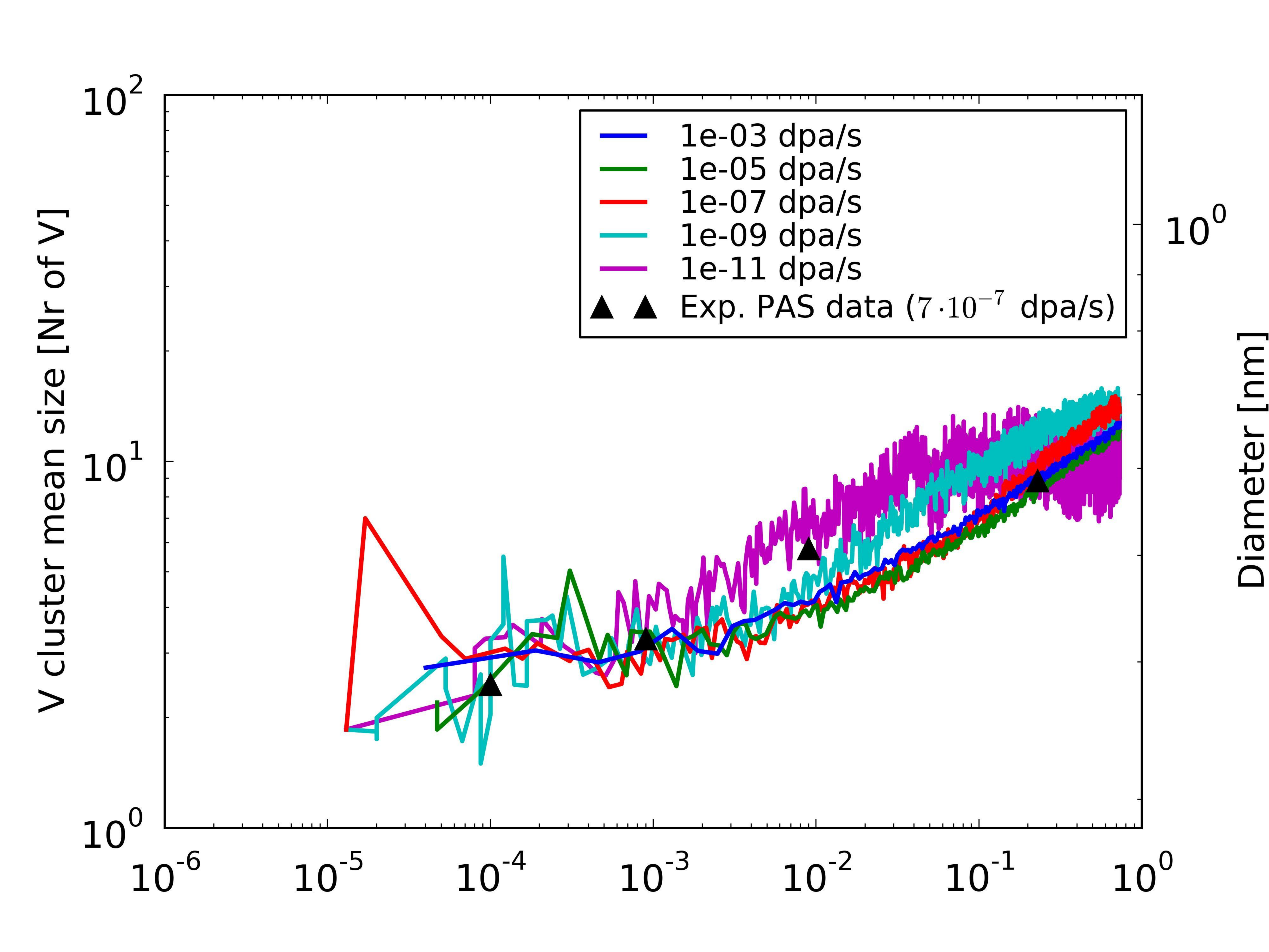}
    \caption{Sensitivity of the results to the dose rate: vacancy cluster mean size versus dpa. The experimental data are
from \cite{eldrup2002dose}.}
  \label{R20130129-a1_vac_mean_size_evolution.pdf}
\end{figure}

The results for the vacancy and visible SIA cluster density evolution are shown in
Fig.~\ref{R20111202-1_vac_dens.pdf} and
\ref{R20120520-00_SIA_dens.pdf}, respectively. It is observed that the density of
both the vacancy and the visible SIA clusters in general are lower with lower flux. The mean size evolution for vacancy and SIA clusters are shown versus dpa in Figs. \ref{R20130129-a1_sia_mean_size_evolution.pdf} and \ref{R20130129-a1_vac_mean_size_evolution.pdf}, respectively. The general trend observed is that lower dose rates increase the growth of clusters, at the expenses of density.

\section{Discussion}\label{sec:discussion}

The effect of the trap concentration, which translates the C content, on the evolution of the density of both vacancy and visible SIA clusters, is found to be moderate. An increase of two orders of magnitude in
the trap density, from 1 appm to 100 appm, only increases the vacancy cluster density by less than one order of magnitude after 0.2 dpa (\textit{Cf}. Fig. \ref{R20120905-0_vac_dens.pdf}). Moreover, there is clearly a saturation of the effect that is quite quickly reached above 50 appm, because at that point many traps remain unused. The overall effect is even less for visible SIA clusters, where no significant change is observed when varying the trap density between 5 appm and 200 appm. To see a significant decrease in cluster density, less than 5 appm traps need to be used. 

The SIA cluster mean sizes are overestimated by about half an order of magnitude, which is a general problem for our model. However, the experimental data by Zinkle and Singh \cite{zinkle2006microstructure} report suprisingly small SIA cluster sizes. Their reported TEM resolution is 0.5 nm for SIA clusters, which is much smaller than the more common value of 1.5 nm  (e.g. in \cite{hernandez2010transmission} the minimum size reported in the given size distributions is 2 nm). We chose to use the latter value as the lower size limit of visible SIA clusters, as we could not see much difference in our results with either value. Mean sizes of visible clusters of 1 nm are, as a matter of fact, never reported: with the exception of this experiment, the smallest mean size found to be reported in the literature is 3--4 nm \cite{malerba2011review}. We therefore suspect that the sizes reported in \cite{zinkle2006microstructure} might be an underestimation, maybe they are in fact the smallest sizes observed.  At any rate, even though the SIA cluster mean sizes are systematically too large, the trends generally agree with the experimental data in all sensitivity studies reported in this paper.

Traps replace in our model the effect of C and CV$_2$ complexes and we have used
in  \cite{jansson2013simulation} 100 appm traps, corresponding to the reported
amount of carbon and nitrogen in the material of the reference experiment
\cite{zinkle2006microstructure,eldrup2002dose}. The results from Sec.
\ref{sec:C_density} show, therefore, that the exact knowledge of the amount of C in the matrix is not crucial to model the cluster density evolution: It can vary even up to 50 \% without any significant change
of the end result. So, at least at this irradiation temperature, variations in the actual C content (in the matrix) are not expected to have any strong influence, so long as the concentration is sufficiently high. Our results also show that, in order to remove the effect of C, it is necessary to reduce its concentration very significantly, down to a level of purity rarely reached in iron.

Our results in \cite{jansson2013simulation} revealed that the fundamental ingredient to reproduce experimental data correctly is that large SIA clusters should be trapped more strongly than small ones. This ingredient makes sure that larger clusters can grow by absorbing more easily de-trapped small clusters, thereby reaching sizes comparable with the experimental ones, and without in the meantime coalescing into a single large cluster. The latter situation would be produced sooner or later if only one trapping energy was used for all sizes \cite{lee2009kinetic}. In our model this size dependence is introduced using the threshold, $N_{th}$. 

Physically, the idea is that large SIA clusters are more likely to interact strongly with the CV$_2$ complexes, as strong interactions only occur with the centre of the cluster. Small SIA clusters are more likely to interact via the edge, in which case the vacancies recombine and only a single C atoms interacts with the SIA cluster, which corresponds to a binding energy of 0.6 eV, independently of the size. The $N_{th}$ parameter fine-tunes these probabilities in a rather rough way, \textit{i.e.} by introducing a step-like function when one should expect a gradual transition. Nonetheless, this approximation proves satisfactory. We see in Sec. \ref{sec:threshold} that a higher value of $N_{th}$ gives a lower density of visible SIA clusters. 

In \cite{jansson2013simulation}, $N_{th}=29$ was fitted to the experimental values for the experiment material in \cite{zinkle2006microstructure}. A value in this range is actually not meaningless if considering the cross-sections for the interactions of C atoms with the centre or the edge of the SIA clusters. SIA clusters are made up of dumbbells in a hexagonal configuration and SIA clusters with a certain number of SIA will form a perfectly symmetric hexagon: $N^i =$ 7, 19, 37$\ldots$ These numbers can be derived from the formula 
\begin{equation}
  N^i_j = 3j(j+1)+1, \quad j \in \mathbb{N}_0.
\end{equation}
The number of edge SIA would be $6j$ for a cluster with $j$ layers of SIA around the central SIA. From this formula it can be seen that a cluster with half of the SIA at the edge has a size somewhere between 19 and 37. A C atom or C-V cluster interacting with a SIA cluster with more than 37 SIA, \textit{i.e.} $N^i>37$, will be more likely to interact with the centre, than with the edge of the cluster. For the dominating C-V clusters at 343 K, this means that they are likely to be bound with a strong binding energy of $\sim$1.4 eV, according to MD calculations \cite{anento2013carbon}. For smaller clusters, the C-V complexes are most likely to interact with the edge and thus with a weak binding energy, 0.6--0.7 eV \cite{anento2013carbon}.

The actual value of the trapping energy for SIA clusters above the threshold size, $N_{th}$, is not important, as long as it is higher than $E^i_t=1.2$  eV (at the irradiation temperature of 343 K). This happens to be, according to atomistic calculation, the order of the binding energy of the CV$_2$ complexes with the centre of the SIA cluster \cite{anento2013carbon}. If the actual energy was higher, the effect would be the same because what is needed is that large SIA clusters, when trapped, must have a sufficiently small rate of de-trapping to be able to grow. Lower values, however, would not have the same effect. If the trapping energy is set too low, no visible SIA clusters form, as
happens with $E^i_t=0.8$ eV.

%

Finally, we have explored the effect of dose rate, which is an important parameter a priori in order to know whether the nanostructural evolution under irradiation leading to embrittlement in the vessel of a power reactor would be the same as in a high flux materials test reactor (or under ion irradiation). The results show that there is a clear dose rate effect on both the vacancy and the visible SIA cluster density evolution, at least at the temperature considered here, which is much lower than the operation temperature of RPV steels. However, a change of the dose rate by eight orders of magnitude only gives a difference of one order of magnitude for both the vacancy and the visible SIA densities, so the effect can be considered limited. It can be summarized by saying that, the lower the dose rate, the more time defects have to cluster, also by coalescence, before new defects are nucleated; or to disappear at sinks before new defects come to make them grow. The overall effect is that the density decreases, while the size increases. The gap to decide whether this will have or not an effect also on mechanical property 
changes is wide. However, one can speculate that the effect on radiation hardening of the increased size with lower dose rate might be offset by decreased density of obstacles to dislocation motion, with overall limited effect on not only hardening, but also embrittlement.

Another important environmental parameter is, of course, irradiation temperature. This, however, changes also the nature of the damage, as the loops observed tend to have $\langle100\rangle$ Burgers vector with increasing temperature. This requires a revision of the model, which is the topic of another paper \cite{jansson2013nanostructure}.

\section{Conclusions}

In conclusion, the present sensitivity study shows that a model for nanostructural evolution under irradiation in iron based on traps translating the effect of C is effective and physically solid, as the variation of the key parameters lends itself to physically consistent interpretations. Moreover, the only fitting parameter used allows the reproduction of experimental results by taking physically reasonable values. Finally, the model allows an assessment of the dose-rate effect that points to a relatively weak one, although the gap between the current model and the embrittlement of RPV steels is too wide to be able to make any definitive conclusion.


\section*{Acknowledgement}

This work was carried out as part of the PERFORM60 project of the 7th Euratom
Framework Programme, partially supported by the European Commission, Grant
agreement number FP7-232612. The authors wish to thank N. Anento, C. S. Becquart,
A. De Backer, C. Domain, A. Serra, and D. Terentyev for advice, assistance and
fruitful discussions during the performance of this work.




\bibliographystyle{model1a-num-names}
\bibliography{/data/people/vjansson/Beam/Articles/vjansson.bib,/data/people/vjansson/Beam/Articles/vjansson_publications.bib}







\end{document}